\documentclass[aps,twocolumn,superscriptaddress,footinbib,longbibliography,prb]{revtex4-1} 

\usepackage{longtable}
\usepackage{morefloats}
\usepackage[dvips]{graphicx}
\usepackage{color}
\usepackage{epsfig,graphicx,amsfonts,amsbsy}
\usepackage{amsmath,amsfonts,amsthm,amssymb}
\usepackage{appendix}
\usepackage{makeidx}
\usepackage{url}
\usepackage{verbatim}
\usepackage[bookmarksnumbered,pdfpagelabels=true,plainpages=false,colorlinks=true,linkcolor=blue,citecolor=blue,urlcolor=blue]{hyperref}
\usepackage[rightcaption]{sidecap}
\usepackage{array}
\usepackage{multirow}
\usepackage{tabularx}
\usepackage{braket} 
\usepackage{dsfont}
\usepackage{bbold}
\usepackage{etoolbox}

\usepackage{dcolumn}
\usepackage{bm}
\usepackage{blindtext}
\usepackage{comment}
\usepackage{physics}
\def\beq{\begin{equation}}
\def\eeq{\end{equation}}
\def\bea{\begin{eqnarray}}
\def\eea{\end{eqnarray}}

\DeclareSymbolFont{bbold}{U}{bbold}{m}{n}
\DeclareSymbolFontAlphabet{\mathbbold}{bbold}

\newcommand{\bs}[1]{\boldsymbol{#1}}

\begin{document}
\title{Topological magnonic dislocations modes}
\author{Carlos Saji}
\affiliation{Departamento de F\'isica, FCFM, Universidad de Chile, Santiago, Chile.}
\author{Nicolas Vidal-Silva}
\affiliation{Departamento de Ciencias F\'isicas, Universidad de La Frontera, Casilla 54-D, Temuco, Chile}
\author{Alvaro S. Nunez}
\affiliation{Departamento de F\'isica, FCFM, Universidad de Chile, Santiago, Chile.}
\author{Roberto E. Troncoso}
\affiliation{Departamento de Física, Facultad de Ciencias, Universidad de Tarapacá, Casilla 7-D, Arica, Chile}
%

\begin{abstract}
Spin fluctuations in two-dimensional (2D) ferromagnets in the presence of crystalline lattice dislocations are investigated. We show the existence of topologically protected non-propagative modes that localize at dislocations. These in-gap states, coined as {\it magnonic dislocation modes}, are characterized by the $Z_2$ topological invariant that derives from parity symmetry broken induced by sublattice magnetic anisotropy. We uncover that bulk topology existing in the perfect crystal is robust under the influence of lattice defects, which is monitored by the real-space Bott index. It is also revealed that the topology of {magnonic dislocation modes} remains unaffected when bulk topology becomes trivial and is remarkably resilient against magnetic disorder. Our findings point to the intriguing relationship between topological lattice defects and the spectrum of topological spin excitations. 
\end{abstract}

\maketitle

\section{Introduction}
Lattice dislocations, structural defects in ordered solids, are irregularities that emerge as abrupt changes in the crystal order. These are characterized by the Burgers vector ${\bf B}$ that remains constant over the entire length of the dislocation. Dislocations in solid-state materials have been the ground for a broad range of scopes such as melting \cite{Nelson1979}, elastic response, and thermal conductivity \cite{friedel1964dislocations}. A renewed interest in dislocations promoted it to a central role in the interplay of real space topological defects and emergent band topology \cite{Ran2009,Lin2023}. Concretely, it was shown in topological insulators and superconductors \cite{Ran2009,Teo2010,Imura2011,Asahi2012,Mesaros2013,Teo2013,Bi2014,Benalcazar2014,Parente2014,Slager2014,Chernodub2017,Panigrahi2022,Schindler2022,Yamada2022}, mechanical \cite{Paulose2015,Xue2021,Ye2022,Deng2022}, and light-based systems \cite{Li2018,Lu2021,Xie2022,AgarwalaPRL2017}, that pair of gapless helical modes appear bound to the line defect, which are determined from the index of the dislocation modes \cite{Ran2009},  
\begin{align}\label{eq: Z2index}
\text{N}_{\mathrm{dis}}=\frac{1}{2\pi} {{\bf B}} \cdot {\bf G} \ (\bmod\, 2)
\end{align}
being the topology of these modes protected by the topological $\mathbb{Z}_{2}$-invariant, ${\bf G}= \nu_{1}{\bf b}_{1}+\nu_{2}{\bf b}_{2}$, where $\nu_{i}$ and ${\bf b}_i$ the weak topological index and reciprocal lattice vectors, respectively. Importantly, these states are robust against disorder that preserve the nontrivial bulk topology, following from the bulk-dislocation correspondence \cite{Geier2021,Liu2021,Kubota2021}, a remarkable feature that has been experimentally demonstrated in 2D photonic crystals and metamaterials.

Quantum spin fluctuations of ordered magnets, magnons, inherit fundamental properties from the crystal lattice structure \cite{Brinkman1967,birss1964symmetry}. It is encoded in their band spectrum and corresponding interaction with phonons, point-like defects, and structural disorder. The role of topological lattice defects on the spin-wave fluctuations has been a recurrent issue, particularly on interference and scattering effects \cite{Pokrovskii1970,Kuchko1998,Turski2009,Gestrin2012}, FMR spectra \cite{Baryakhtar1968,Zmijan1980,Schmidt1988}, relaxation \cite{Morkowski1968,Morkowski1974} and thermal conductivity \cite{Baryakhtar1967,Fomethe1982}, and recently in helical textures on chiral magnets \cite{AzharPRL2022}. Differently, the concept of topology might emerge in the band structure of magnonic states with remarkable signatures such as robust helical edge-states and thermal Hall effect \cite{Li2021,Zhuo2023}. Topological magnons have been strongly scrutinized in a wide variety of spin and lattice systems \cite{TM3,TM5,TM7,TM13,TM15,Wang2021,Li2021,Zhuo2023}, which are characterized by topological invariants that remain unchanged under smooth deformations and set the ground for the bulk-boundary correspondence \cite{mm}. The immunity of topological states to disorder, deep-rooted to topological matter \cite{TI1,TI2}, has been tested in collinear magnets \cite{WangPRL2020,Akagi2020} and glassy skyrmions \cite{rosales2024robustness}. However, the influence of crystal lattice defects, such as dislocations, on the band topology is an unexplored arena in magnetic systems with intriguing effects regarding the stability and localization of topological magnon states. 

\begin{figure}[thb]
\includegraphics[width=\columnwidth]{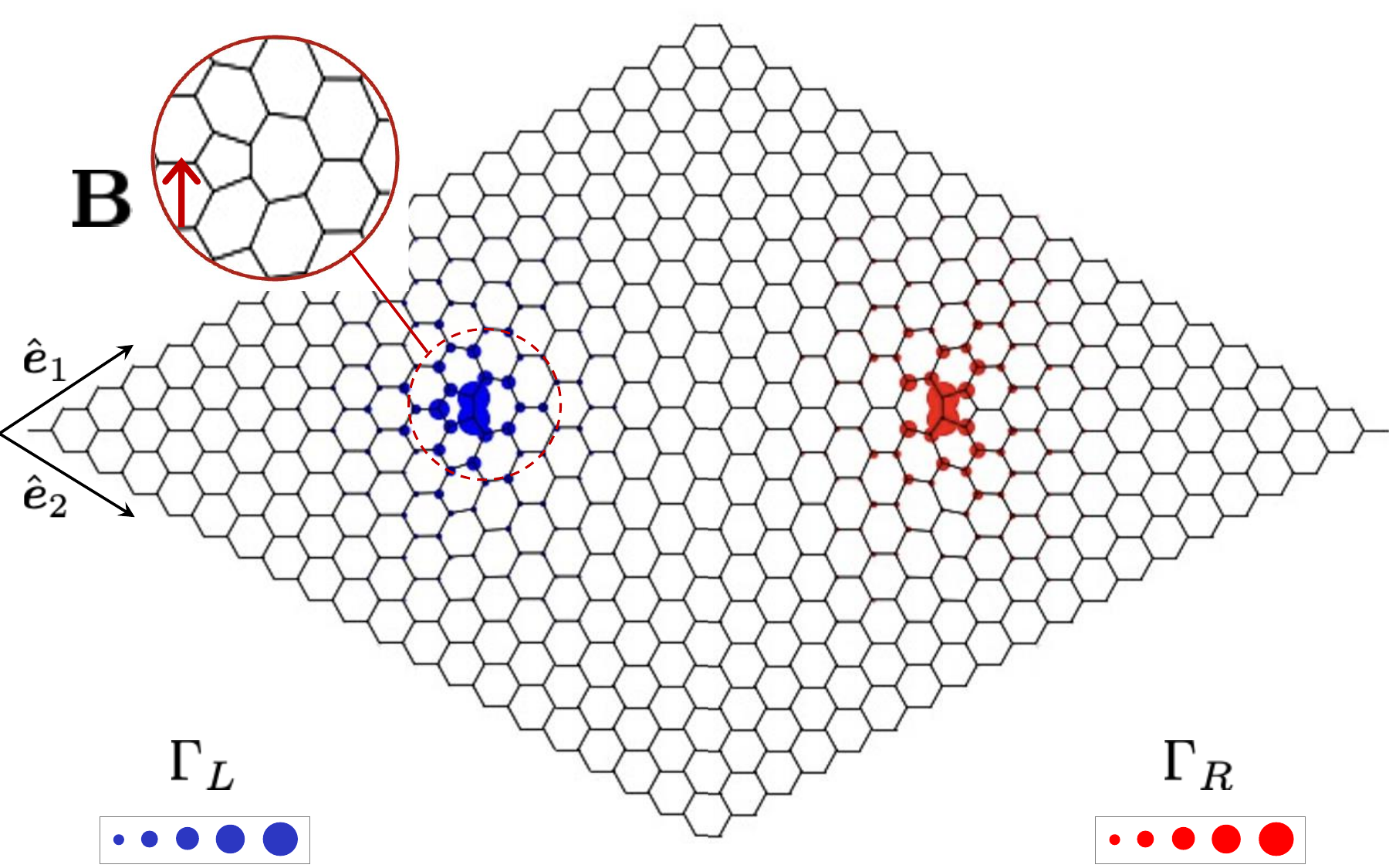}
\caption{Schematic representation of a pair of dislocations in the hexagonal lattice with the in-plane burgers vector ${\bf B}=\pm \sqrt{3}\bs{ \hat{y}}$, at the left and right side of the dislocations, respectively. The topological {magnonic dislocation modes}, $\Gamma_L$ (left) and $\Gamma_R$ (right) states, are shown localized at the ends of the dislocation. These modes are gapped due to parity symmetry breaking, and their wave function amplitudes are, accordingly, colored differently.}
\label{fig1}
\end{figure}

In this paper, we show the existence of topological magnon states bound to the dislocations (see Fig. \ref{fig1}) within the crystal structure of magnetic insulators. Remarkably, it is shown that {magnonic dislocation modes} are stabilized when parity symmetry is broken and persists while the topological bulk gap is closed. The existing bulk topology is diagnosed by the evaluation of a real-space topological index, the bosonic Bott index, which turns out to be stable in the presence of dislocations. It is shown that { dislocation modes} and edge-modes have different degrees of robustness against disorder. The stability of these states is demonstrated for a model that supports a collinear ferromagnetic phase.

\section{Spin and lattice model}
We consider a magnetic system with spins localized on a two-dimensional honeycomb lattice described by the spin Hamiltonian,
\begin{align}
\mathcal{H}_S=-\sum_{\langle \bs r\bs r'\rangle}&\nonumber\left[J{\bs S}_{\bs r}\cdot{\bs S}_{\bs r'}+F
\left({\bs S}_{\bs r}\cdot\mathbf{e}_{\bs r\bs r'}\right)\left({\bs S}_{\bs r'}\cdot\mathbf{e}_{\bs r\bs r'}\right)\right.\\
&\left.\qquad+\left(K_{\bs r}(S^z_{\bs r})^2
-B S^z_{\bs r}\right)\delta_{\bs r\bs r'}\right],
\label{eq: spin-hamiltonian}
\end{align}
with the nearest-neighbor exchange coupling $J$ and pseudodipolar interaction with strength $F$. The last coupling results from the spin-orbit interaction \cite{JackeliPRL2009,Wang2017}, being $\mathbf{e}_{\bs r\bs r'}$ the unit vector that connects $\bs r$ and $\bs r'$ lattice sites. The easy-axis anisotropy, $K_{\bs r}$, at the ${\cal A}({\cal B})$-sublattice is parametrized as $K_{{\cal A}({\cal B})}=K\pm\Delta_K$, and
$B$ is the applied magnetic field along the $z$-direction. The dislocation is introduced following the Volterra process \cite{Kleinert}. A set of atoms are removed, resulting in a pentagon-{heptagon} structure, and thus, moving an equal number of steps around the dislocation sets a non-closed loop defining the Burgers vector, see Fig. \ref{fig1}. {It is worth to stress that, although linear dislocations might be more complicated, e.g., curvilinear or multiples, we choose the displayed linear straight dislocation for simplicity.} For numerical convenience, we consider a couple of defects that create a pair of dislocations in the bulk of the system. The local lattice deformation produces a distortion of the spin Hamiltonian, $\mathcal{H}_{T}=\mathcal{H}_S+\mathcal{H}_{\text{dis}}$, that gives rise to a magnetoelastic interaction $\mathcal{H}_{\text{dis}}$ relating spin and lattice degrees of freedom. { Hereafter, in our numerical calculations, we consider the relevant magnetic parameters with similar order of magnitude, i.e., $J\sim K\sim \Delta_K$ as typically occurs in Van der Waals magnets \cite{kim2019exploitable,jiang2018spin}}

Quantum spin fluctuations around the ordered ground state are determined by the Hamiltonian of non-interacting magnons using Holstein-Primakoff (HP) bosons \cite{HolsteinPR1940}. Around the classical magnetic state, {$z$-axis oriented}, the HP mapping of spin operators reads ${S}^{+}_{\bs r}=\left(2S-a^{\dagger}_{\bs r}a_{\bs r}\right)^{1/2}a_{\bs r}$, ${S}^{-}_{\bs r}=\left(2S-a^{\dagger}_{\bs r}a_{\bs r}\right)^{1/2}a^{\dagger}_{\bs r}$, and ${S}^{z}_{\bs r}=S-a^{\dagger}_{\bs r}a_{\bs r}$. Thus, expanding the spin operators as a series in $1/S$, the spin Hamiltonian reduces to ${\cal H}_T\approx {\cal H}_0+{\cal H}_m$, being ${\cal H}_0$ the classical and zero-point energy. In real space the magnon Hamiltonian reads ${\cal H}_m=\Psi^{\dagger}{\text H}\Psi$, { where the $2N$-components operator field $\Psi=\left(a_{\bs r_1}\dots,a_{\bs r_N},a^{\dagger}_{\bs r_1}\dots a^{\dagger}_{\bs r_N}\right)^T$, with $\bs r_i$ the lattice position and ${\text H}$ the $2N\times 2N$ matrix Hamiltonian, with $N$ the number of lattice sites and tight binding matrix elements given by:
\begin{align*}
{\text H}_{ \bs r\bs r'}=\left(\begin{array}{cc}
\Omega_{ \bs r\bs r'} & \Delta_{ \bs r\bs r'} \\
\Delta_{ \bs r\bs r'}^{*} & \Omega_{ \bs r\bs r'}^{*}
\end{array}\right)
\end{align*}
where $\Delta_{ \bs r\bs r'}=-S F (\boldsymbol{e}^{-} \cdot \mathbf{e}_{\bs r\bs r'}) (\mathbf{e}_{\bs r\bs r'} \cdot  \boldsymbol{e}^{-}) \delta_{\langle \bs r\bs r'\rangle}/2$, $\Omega_{ \bs r\bs r'}=-S\left[J + F (\boldsymbol{e}^{+} \cdot \mathbf{e}_{\bs r\bs r'}) (\mathbf{e}_{\bs r\bs r'} \cdot  \boldsymbol{e}^{-})\right]\delta_{\langle \bs r\bs r'\rangle}/2 +S\Lambda_{\bs r}\delta_{ \bs r\bs r'}$, $\Lambda_{\bs r}=\sum_{\langle \bs r\bs r'\rangle}\left[J{\bs S}_{\bs r}\cdot{\bs S}_{\bs r'}+ \left(K_{\bs r}(S^z_{\bs r})^2
-B S^z_{\bs r}\right)\delta_{\bs r\bs r'}\right]$, and $\boldsymbol{e}^{\pm}=\bs{ \hat{x}}\pm i \bs{ \hat{y}}$}. \begin{figure}[!htb]
\includegraphics[width=\columnwidth]{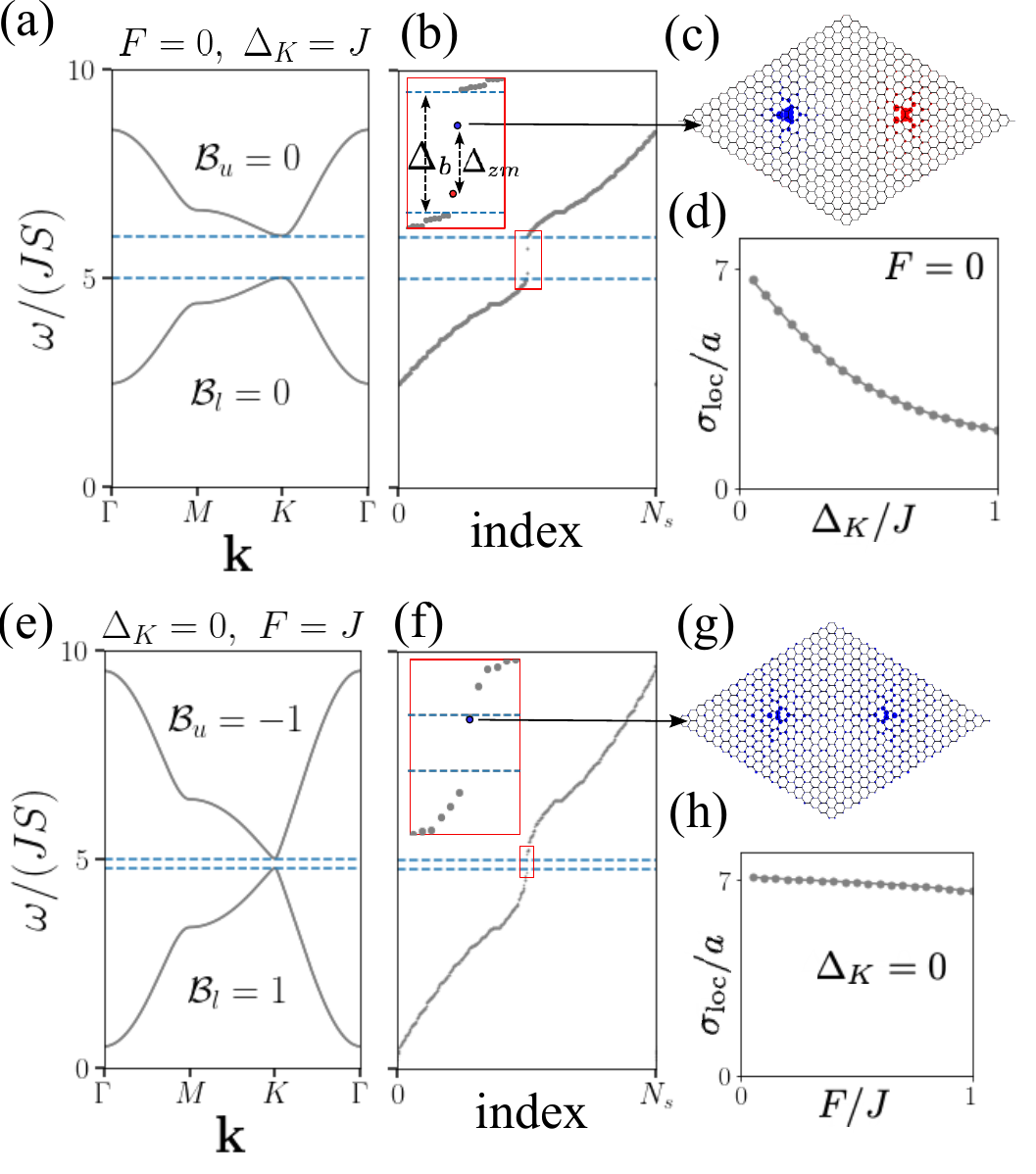}
\caption{{Top panel: Magnon spectrum for $S=1$, $K=2 J$, $F=0$, $\Delta_{K}=J$ and $B=0$. Bottom panel: Magnon spectrum for $S=1$, $K=2 J$, $F=J$, $\Delta_{K}=0$ and $B=0$. (a) and (e) Bulk energy spectrum without dislocation along high-symmetry points in the Brillouin zone. The Bott index for the lower and upper bulk bands satisfy $\mathcal{B}_u=-\mathcal{B}_l=0$ and $\mathcal{B}_u=-\mathcal{B}_l=-1$ at panel (a) and (e), respectively. (b) and (f) magnonic energy spectrum with PBCs in presence of a dislocation pair. The energy of dislocation modes are within the band-gap. (c) and (g) real spatial amplitude of the dislocation modes wave function $\Gamma_{L,R}(\bs r)$. (d) and (h) Real space localization, $\sigma_{\mathrm{loc}}$, of $\Gamma_{L}(\bs r)$ as a function of $\Delta_K/J$ and $F/J$, respectively.}}
\label{fig: spectrum-bott}
\end{figure}
The bosonic Hamiltonian is para-diagonalized by the Bogoliubov transformation $(a_{\bs r},\dots,a^{\dagger}_{\bs r})^T={\text T}_{\bs r\bs r'}(\alpha_{\bs r'},\dots,\alpha^{\dagger}_{\bs r'})^T$, with ${\text T}$ the paraunitary transformation that satisfy ${\text T}^{\dagger}\zeta{\text T}=\zeta$ to guarantee the commutation relation $\left[{\bs \alpha},{\bs \alpha}^{\dagger}\right]=\mathbb{I}\otimes\sigma_z=\zeta$ for bosonic operators \cite{colpa1978diagonalization} {(with $\sigma_z$ the Pauli matrix}). Therefore, the diagonalized magnon Hamiltonian is written as ${\cal H}_{m}=\sum_{n}{\cal E}_{n}\alpha^{\dagger}_{n}\alpha_{n}$ with ${\cal E}_{n}$ the energy for the $n^{\text{th}}$-band.

\section{Bulk topology}
Ferromagnetic honeycomb defects-free lattices, described by the spin Hamiltonian at Eq. (\ref{eq: spin-hamiltonian}) exhibit topological magnonic phases featured by the Chern number \cite{Wang2017,WangPRL2020,Wang2021}.
The topological gap, $\Delta_b$, is induced by the pseudo-dipolar interaction $F$ and controlled by the sublattice easy-axis anisotropy $\Delta_K$. We now determine the bulk topology when dislocations are present. Since crystalline symmetry is locally broken, we evaluate the topology of magnonic bands through the bosonic Bott index \cite{WangPRL2020}. It is a real-space topological invariant that is equivalent to the Chern number in the thermodynamic limit and when translational invariance is restored \cite{Loring_2010,Toniolo2018,Toniolo2022}. For the set of eigenstates $\left\{{\cal E}_n\right\}$, it is defined as ${\cal B}\left({\cal E}_n\right)=\text{Im}\left[\text{Tr}\left[\text{log}\left(V_YV_XV_Y^{\dagger}V^{\dagger}_X\right)\right]\right]/2\pi$, where $V_X$ and $V_Y$ are unitary matrices defined by
\begin{gather}
Pe^{i\pi \Theta}P={\text T}\zeta \left(\begin{matrix}0 & 0 \\0 & V_\Theta \end{matrix}\right){\text T}^\dagger\zeta,
\end{gather}
with $\Theta=X,Y$ the position operators, {represented by matrices where elements are normalized lattice position}. The projector $P={\text T}\zeta\Gamma_{\cal N}{\text T}^\dagger\zeta$ on states $\left\{{\cal E}_n\right\}$ and the diagonal matrix $[\Gamma_{\cal N}]_{nn'}=\gamma\delta_{nn'}$, with $\gamma=0$ for ${\cal N}<n$, and $\gamma=1$ when $1\leq n\leq{\cal N}$ \footnote{For fermionic systems the metric $\eta=\mathbb{I}$ and the definition in the main text returns to the electronic Bott index}. For clean systems, the Bott index of each band is well defined and is an integer as long as $V_YV_XV_Y^{\dagger}V^{\dagger}_X$ is nonsingular. In particular, $\mathcal{B}=0$ when $V_X$ and $V_Y$ commute and the corresponding band is topologically trivial.

We now consider the effects of dislocations on the band structure and topology of magnonic states. First, in the perfect hexagonal crystal, the two-band spectrum of topologically trivial and non-trivial magnon excitations are displayed at Figs. \ref{fig: spectrum-bott}(a) and \ref{fig: spectrum-bott}(e), respectively. Note that the trivial gap, induced at the Dirac point, is the result of breaking the parity symmetry by the sublattice anisotropy difference $\Delta_{K}$. In the presence of dislocations, two interesting effects are highlighted from the magnonic spectrum displayed in Figs. \ref{fig: spectrum-bott}(b) and (f). First, the bulk topology prevails since a non-zero Bott index is found for the top (${\cal B}_u=-1$) and bottom (${\cal B}_l=+1$) bands, for $F>0$ and different dislocations lengths, see panel (f). In particular, the Bott index vanishes when $F=0$ and the bulk topology becomes trivial as expected, see panel (b). Secondly, a pair of gapped {\it {magnonic dislocation modes}} appear inside the gap, $\Delta_b$, and bound to the ends of the dislocation, indicated by the black dots at the inset of Fig. \ref{fig: spectrum-bott}(b). The gap between these states, $\Delta_{dm}$, is induced by non-zero values of $\Delta_{K}$, which is independent of the pseudo-dipolar coupling, and therefore, becomes gapless once the parity symmetry is restored, see inset of Fig. \ref{fig: spectrum-bott}(f). It is worth noting that for a strip geometry, we find that the energy of { dislocation modes} coexist with those of topological edge-states in the presence of pseudo-dipolar energy, where open boundary conditions (OBCs) along the $\hat{e}_1$ axis is assumed, as is shown in Fig. \ref{fig: spectrum-bott}(b).

{Magnonic dislocation modes} are localized at the ends of the dislocation. Their spatial localization, displayed at Fig. \ref{fig: spectrum-bott}, is determined from the overlap, $\Gamma_{n}({\bs r})=\left|\langle GS|a_{\bs r}\alpha^{\dagger}_{n}|GS\rangle\right|^2$, between the local excitation and the eigenstates, with $|{GS}\rangle$ the ground-state of the magnon Hamiltonian. We denote $\Gamma_{L,R}({\bs r})$ to the magnonic contribution of the left- and right { dislocation modes} wave function, displayed in Fig. \ref{fig1} and \ref{fig: spectrum-bott}. 
\begin{figure}[!htb]
\includegraphics[width=\columnwidth]{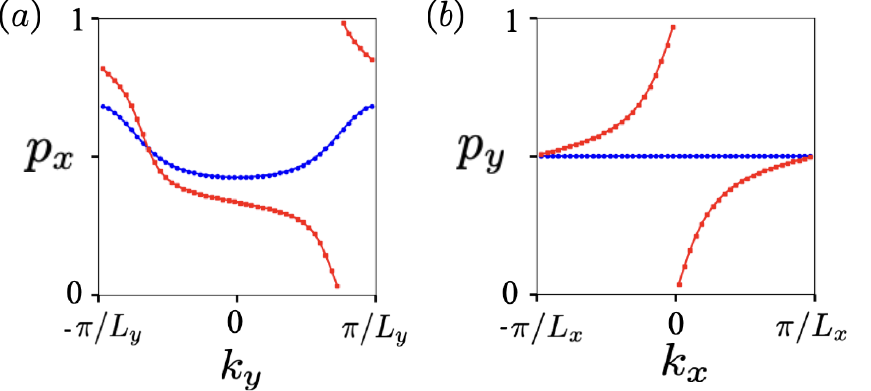}
\caption{{ Left panel (a): polarization $p_{x}(k_{y})$ and right panel (b): polarization $p_{y}(k_{x})$, for the cases with bulk topology trivial $K=2J,\Delta_{K}=J,F=0,B=0$ (blue line), and bulk topology is nontrivial trivial $K=2J,\Delta_{K}=0,F=J,B=0$ (red line). In all plots we assumed $J=1$.}}
\label{fig: wanniercenters}
\end{figure}
The {magnonic dislocation modes} are non-propagating states, strongly bound to the lattice dislocation with localization length $\sigma_{\text{loc}}$. For a strip geometry, assuming finite size along $\hat{\bs e}_1$ direction, the corresponding eigen states for the dislocations modes are displayed at panel (c) and (g). In Fig. \ref{fig: spectrum-bott}(d), we plot the characteristic localization length of the wave function around the dislocation as a function of $\Delta_K$. It is defined $\sigma_{\text{loc}}=\int |\boldsymbol{r}-\boldsymbol{r}_{\text{dis}}||\psi(\boldsymbol{r})|^{2} \ d^{2}r$, {where we choose $\psi(\boldsymbol{r})=\Gamma_{L,R}(\boldsymbol{r})$ for the magnonic wavefunction} and   $\boldsymbol{r}_{\text{dis}}$ the position of the dislocation. Interestingly, the magnonic zero-mode wave function tends to delocalize as $\Delta_K$ becomes null, corresponding to the  {magnonic dislocation modes} becoming gapless, see panels (d) and (h). {Importantly, we remark that while the pseudo-dipolar interactions play a critical role in stabilizing the bulk topological modes, it is not necessary in a minimal model to give rise topological dislocation magnon modes as long as the sublattice-breaking anisotropy is present (see Fig. \ref{fig: spectrum-bott}(c)). The pseudo-dipolar interaction introduce additional symmetry-breaking terms that strength the stability of the dislocation modes. 
}

\section{Weak topology and $\mathbb{Z}_{2}$-invariant}
We now establish the topological properties of {magnon modes} bound to the lattice dislocations. The topology of the {magnonic dislocation modes} is determined by the $\mathbb{Z}_{2}$ topological index $\bf{G}$ which, through Eq. (\ref{eq: Z2index}), determines the number of modes appearing at the dislocation. The $\mathbb{Z}_{2}$-invariant and weak topological indices, $\nu_{1}$ and $\nu_{2}$, are found through the bulk polarizations, defined as the sum of the Wannier centers,
\begin{align}
p_{x}(k_{y})&= \sum_j \nu_x^j\left(k_y\right)  \pmod{1},\\
p_{y}(k_{x})&= \sum_j \nu_y^j\left(k_x\right)  \pmod{1},
\end{align}
which are determined by diagonalizing the Wilson loop matrix, $W_{\boldsymbol{k}}|\nu_{\boldsymbol{k}}^j\rangle=e^{2\pi i \nu_x^j\left(k_y\right)}|\nu_{\boldsymbol{k}}^j\rangle$, defined by the product
\begin{align}
\mathcal{W}_{k_x}=F_{k_x+\left(N_x-1\right) \Delta k_x, k_y} \cdots F_{k_x+\Delta k_x, k_y} F_{k_x, k_y},
\end{align}
where $\left[F_{k}\right]^{m n}=\left\langle u_{k_{x}+\Delta k_{x},k_{y}}^m \mid u_{k_{x},k_{y}}^n\right\rangle_{\mathrm{para}}$ are the overlap thought the discrete path $k_{j}=k_{x}+j\Delta k_{x}$. { The weak topological nature of the system is determined by the winding of the Wilson loops along certain reciprocal lattice directions. The Wilson loop, calculated as the phase evolution of eigenstates over a closed trajectory in momentum space, provides a measure of the actual topological invariant. For dislocation-bound modes, the weak topological invariant, derived from the Wilson loop, ensures the existence of states localized at the dislocation. These states arise because the dislocation {\it cuts through} the crystal, exposing a one-dimensional topological invariant encoded in the bulk. This mechanism is distinct from strong topology, which relies on the presence of robust edge-states in 2D systems.} Here, we employ this formulation to compute numerically the Wannier centers $\nu^{j}_{x,y}$  and, therefore, the polarizations $p_x(k_y)$ and $p_y(k_x)$. In Fig. \ref{fig: wanniercenters}, we show the polarizations, $p_{x,y}$, { evaluated for two scenarios: $K=2J,\Delta_{K}=0,F=1,B=0$, where bulk topology is nontrivial ($F>0$) and parity symmetry is preserved ($\Delta_K=0$); and $K=2J,\Delta_{K}=J,F=0,B=0$  where bulk topology is trivial ($F=0$) and parity symmetry is broken ($\Delta_K>0$)}. In all plots we assumed $J=1$.  

On the other hand, the polarization is related to the Zak phase as follows $p_{\mu}(k_{x,y})=\frac{i}{2\pi}\int \sum_{n} \mathcal{A}_{n,\mu}(\bs k)dk_{y,x}$, where ${\mathcal{A}_{n,\mu}}(\boldsymbol{k})= \mathrm{Tr}[\Gamma_{n}\Sigma_{z}T_{\boldsymbol{k}}^{\dagger}\Sigma_{z}\partial_{k_{\mu}}T_{\boldsymbol{k}}]$ is the Berry connection, {with $\Sigma_{z}=\sigma_0 \otimes \sigma_z$},  and the summation runs over the lower bands (below the gap). A quantized polarization indicates that the system lies in a topological phase and provides information about high-order topological magnonic states \cite{LiNPJ2019}. Moreover, the weak $\mathbb{Z}_{2}$ invariants in Eq. (\ref{eq: Z2index}) can also be computed through the Zak phase \cite{Teo2013},
\begin{align}
\nu_{\mu}=\frac{i}{\pi} \int_{\mathcal{C}_{\mu}} \sum_{n}\mathcal{A}_{n}(\bs k) \cdot d \bs k,    
\end{align}
where the $1$-cycles $\mathcal{C}_x=\left.B Z\right|_{k_x=\pi/L_{x}}$  and $\mathcal{C}_y=\left.B Z\right|_{k_y=\pi/L_{y}}$  run along the $k_x$ and $k_y$ directions in the Brilloin zone, respectively.  In particular,  we have that $\nu_{x,y}=2p_{x,y}(k_{y,x}=\pi/L_{y,x})$, and according to the previous numerical calculations (see Fig. \ref{fig: wanniercenters}), we deduce that $(\nu_{x},\nu_{y})=(0,1)$, and hence the $\mathbb{Z}_{2}$-weak invariant is ${\bf G}= \boldsymbol{b}_{y}= \frac{2\pi}{\sqrt{3}}\boldsymbol{\hat{y}}$. On the other hand, the Burger vector is given by ${\bf B}=q(\boldsymbol{a}_{2}-\boldsymbol{a}_{1})=q\sqrt{3}\boldsymbol{\hat{y}}$, where $q=\pm 1$ is the charge of the dislocation (see Fig. \ref{fig1}). Finally, we arrive at $\text{N}_{\mathrm{dis}}=\pm 1$, is non-trivial. Therefore, there must be a topologically protected dislocation mode, as we claimed.

\begin{figure}[!htb]
\centering
\includegraphics[width=0.45\textwidth]{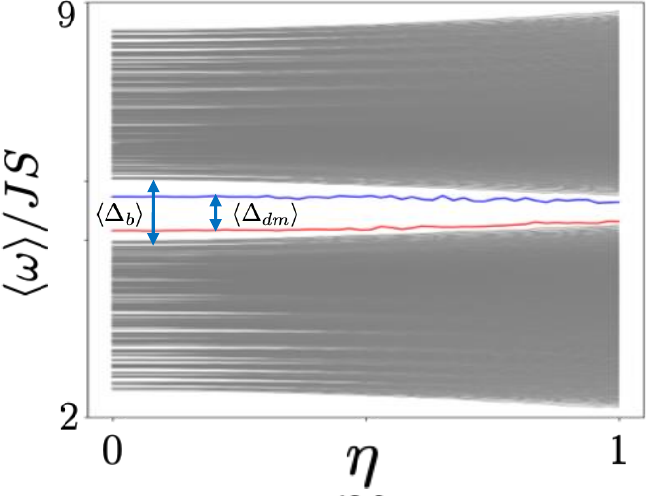}
\caption{The energy spectrum of magnons as a function of disorder strength $\eta$ averaged over $20$ realizations { with parameters $S=1,K=2J, F=0,\Delta_{K}=J$}. {Energy gaps associated with both bulk and {magnonic dislocation modes} are depicted in blue and red lines.}}
\label{fig:spectrum_disorder}
\end{figure} 

\section{Robustness against disorder} Thermal fluctuations, noise, and disorder are ubiquitous and might cause negative effects on the robustness of topological properties. We now discuss the stability of existing topologically protected {magnon modes} at the dislocation against magnetic disorder. The disorder is modeled by a random out-of-plane magnetic field across the sample, $\mathcal{H}_{\mathrm{random}}=\sum_{i}\chi S^{z}_{i}$ where $\chi \in [-\eta,\eta]$ is a random number and $\eta$ is the disorder strength. 
{Although a general models for disorder would provide a more realistic representation of imperfections, the choice of $\mathcal{H}_{\mathrm{random}}$ is made for its simplicity and to capture robustness effects of topological modes. A systematic analysis of disorder is left for future studies.} Disorder-averaged magnon spectrum as a function of disorder strength $\eta$ is depicted in Fig. \ref{fig:spectrum_disorder}. The results are averaged over $n=20$ realizations of disorder in the spin lattice, where we set the anisotropy and the sublattice anisotropy difference at { $K=2J$ and $\Delta_K=J$}, respectively. {Magnonic dislocation modes} are remarkably robust against the effect of disordered magnetic impurities with considerable strength, resulting from topological protection. The energy of these states remains isolated within the magnon gap $\Delta_b$, avoiding hybridization with bulk states. Although translational symmetry is broken by the presence of disorder, the localization of magnon modes at the dislocation is not disrupted and their spatial distribution prevail.

\section{Discussion and conclusions}
The experimental realization of topological {magnonic dislocation modes} might settle in two stages: first, the control of the geometrical properties of dislocations and second, on the actual excitation and local detection of different magnonic states. Advances in manufacturing and imaging techniques allow for a feasible control of dislocations density \cite{zheng2015frontiers,yu2015situ} on the lattice, where various techniques such as XTM, STM, and AFM would enable the real-time observation of dislocations \cite{zheng2015frontiers,yu2015situ,callahan2018direct,schaff2001situ,tsuji2001observation}. {  For instance, dislocations in van der Waals magnets (e.g., CrI$_3$ or Fe$_3$GeTe$_2$) could be engineered through techniques such as focused ion beam milling \cite{telkhozhayeva2024roadmap} or controlled crystal growth that introduces and controls topological lattice defects \cite{pan2022growth,meisenheimer2023ordering}. In addition, the control of dislocations density over such materials can be engineered by external strains \cite{qi2023recent,yu2024manipulating}}. {Magnonic dislocation modes}, as well as other states, can be excited by time-dependent (RF) magnetic fields, where their detection could be achieved by quantum metrology techniques \cite{andrich2017long,purser2020spinwave}, such as nitrogen-vacancy (NV) centers, which are spin sensors that provides a local monitoring of the spatial localization of the wave function near the dislocations. Matching with the frequency of other magnonic states,  would allow to detect their coexistence with topological edges-states.

{ Although some features of the dislocation mode, such as its localized nature and topological protection, are reminiscent of phenomena in electronic systems, the magnonic system exhibits distinct physics due to their bosonic nature. Magnons are non-conserved quasiparticles whose lifetime and thermal population depend on external driving and dissipation mechanisms, leading to non-equilibrium dynamics of the topological modes that are fundamentally different from their electronic counterparts. Also, the localized nature of the topological dislocation modes could facilitate potential advantages in spintronics, as a magnonic emitter where the mode serves as a controllable source of spin waves. Moreover, the topological protection of the mode may allow for robust, low-dissipation energy transport, which is particularly desirable for magnon-based devices. Lastly, the magnon spectrum is highly tunable via external magnetic fields, mechanical strains, and temperature, enabling versatile control of the topological properties.}

In summary, we have shown that bulk topology prevails in the presence of linear topological defects and dislocations in 2D hexagonal lattices. In addition, such defects induce magnonic states bound at the ends of dislocations. These states are topologically protected and classified by the $\mathbb{Z}_{2}$-invariant, stabilized by the breaking of parity symmetry and existing even for trivial bulk topology. At the dislocation, the pair of gapped {localized topological} modes are determined by the relation between the Burgers vector and the topological $\mathbb{Z}_{2}$-invariant. The presented model is general and might be employed to other forms of magnetic order.

\vspace{1em}
\section{Acknowledgments}
C.S. thanks the financial support provided by ANID National Doctoral Scholarship Nº 21210450. R.E.T thanks funding from Fondecyt Regular 1230747. A.S.N acknowledges funding from Fondecyt Regular 1230515. N.V-S. thanks funding from Fondecyt Iniciacion 11220046.

\bibliography{dislocatedtopmag}

\onecolumngrid

\end{document}